# Collision of Environmental Injustice and Sea Level Rise: Assessment of Risk Inequality in Flood-induced Pollutant Dispersion from Toxic Sites in Texas

Zhewei Liu[1], Ali Mostafavi[1]*

[1] UrbanResilience.AI Lab, Zachry Department of Civil and Environmental Engineering, Texas A&M University, College Station, TX, 77843

*mostafavi@tamu.edu

## Abstract:

Global sea-level rise causes increasing threats of coastal flood and subsequent pollutant dispersion. However, there are still few studies on the disparity arising from such threats and the extent to which different communities could be exposed to flood-induced pollution dispersion from toxic sites under future sea level rise. To address this gap, this study selects Texas (a U.S. state with a large number of toxic sites and significant flood hazards) as the study area and investigates impacts of flood-induced pollutant dispersion on different communities under current (2018) and future (2050) flood hazard scenarios. The results show, currently, north coastline in Texas bears higher threats and vulnerable communities (i.e., low income, minorities and unemployed) are disproportionally exposed to these threats. In addition, the future sea-level rise and the exacerbated flood hazards will put additional threats on more (about 10%) Texas residents, among which vulnerable communities will still be disproportionately exposed to the increased threats. Our study reveals the facts that potential coastal pollutant dispersion will further aggravate the environmental injustice issues at the intersection of toxic sites and flood hazards for vulnerable populations and exacerbate risk inequalities. Given the dire impacts of flood-induced pollution dispersion on communities' health, the findings have important implications for specific actions from the policy makers to mitigate the inequitable risks.

# 1. Introduction

About 60% of non-federal National Priorities List (NPL) sites (i.e., the priority list of hazardous waste sites) in the United States, are in the areas that may be affected by impacted by the climate change effects (Gómez 2021). The projected sea level rise and the accompanying coastal flood can cause pollutants dispersion from the sits and threaten public health. Moreover, the overburdened communities of color and low-income may disproportionally dwell in the regions around these sites and bear particular health impacts (Crawford 1994, Carter and Kalman 2020, Huang, Lu et al. 2022), which worsens the living conditions of socially vulnerable communities and gives rise to exacerbated environmental injustice issues in the context of global climate change.

Efforts have been made by previous studies to investigate flood risk and its environmental justice implications. Studies have found that socially vulnerable communities from different backgrounds experience drastic inequality in exposure to the hazard risk, and such inequity show varied extent across different regions (Collins, Grineski et al. 2019, Wang, Zhang et al. 2020, Zhan, Shi et al. 2022). For example, in the context of UK, some studies found that no displayed socioeconomic disparity regarding to the risks of riverine/inland pre-flood (e.g., residence in 100-year flood zones), but regions with lower socioeconomic status are exposed to higher coastal risks (Fielding 2007, Walker and Burningham 2011).While studies in the US has documented mixed findings: socially advantageous communities bear more risk towards pre-land risks in some regions such as Miami, while disadvantageous communities bear more risk in other regions, e.g., Houston (Chakraborty, Collins et al. 2014). The built environment and amenities can be possible factors contributing to the difference across different regions (Chakraborty, Collins et al. 2014, Yao, Shi et al. 2021, Zhang, Shi et al. 2022).

However, one important limitation of the existing literature is the insufficient attention to evaluating the intersection of flood exposure and environmental justice issues surrounding toxic sites. In particular, the existing studies have paid limited attention to exacerbated flood hazard exposure of toxic sites under future sea-level rise scenarios. Recognizing this important gap, this study examines the coastal area of Texas, USA, which has a large number of toxic sites and significant flood hazard exposure as a case study region to answer the following research questions:

- RQ1: To what extent coastal regions are threatened by pollutant dispersion from flood under current flood hazard scenario?
- RQ2: What is the extent of hazard exposure disparity among vulnerable populations?
- RQ3: To what extent the future flood hazards exacerbated by sea level rise increase the threats of pollutant dispersion exposure?
- RQ4: How the increase in threats is disproportionately affecting vulnerable populations?

To address these questions, we utilize the socioeconomic datasets, and coastal inundation maps in the current and future scenarios for the analysis. The results reveal that, in the current scenario (2018), the areas along the north coastline in Texas (such as Houston and Beaumont), are subject to higher threats to pollution dispersion, and that the communities of low-income, minority, unemployed suffer from greater threats. Moreover, the future sea level rise by 2050 will increase the flood threats to more people, among whom vulnerable populations will be disproportionately exposed to flood-induced pollutant dispersion from toxic sites.

The remaining sections of this paper is structured as follows: Section 2 summarizes the previous relevant studies; Section 3 elaborates the datasets and methodology; Section 4 details the experimental results;

Section 5 discusses the study implications and proposes practical suggestions for policy makers; Finally, Section 6 discusses the study limitations and future research directions.

## 2. Background

Floods threaten human life in various ways. One notable way is the impacts on human health, including injuries, diseases and psychological trauma (Du, FitzGerald et al. 2010, Stanke, Murray et al. 2012, Zhong, Yang et al. 2018, Graham, White et al. 2019, Palinkas and Wong 2020) . The landscape of the flooded areas can be permanently changed and contaminated, and expose the residents to long-term chronic diseases (Euripidou and Murray 2004, Karaye, Stone et al. 2019), and also damage the properties and built environments (Brody, Zahran et al. 2008, Merz, Kreibich et al. 2010, Alipour, Ahmadalipour et al. 2020, Kousky, Palim et al. 2020).

Environmental justice is another concerned issue associated with flood hazards. Due to the socioeconomic status and cultural background, different demographic communities suffer from unequal flood hazards exposure and impacts. The studies from different regions of the world have shown different findings. Some UK-based studies conclude that people of lower socioeconomic status bear higher risk from coastal flood (Fielding 2007, Walker and Burningham 2011). Studies from US reveal that in Miami, the non-Hispanic Black and Hispanic are disproportionately exposed to inland flood risk, and underrepresented with coastal flood risk (Chakraborty, Collins et al. 2014). And the study from Canada shows that senior people (over 65 years old) are the groups that are more vulnerable to coastal climate change in Atlantic Canada (Manuel, Rapaport et al. 2015). These findings demonstrate the significance of examining the intersection of environmental justice issues and flood hazards, especially under future climate change impacts such as sea level rise.

A peculiar type of flood impacts is pollutant dispersion from industrial and toxic sites that can be inundated due to flooding. Studies from the US show that vulnerable communities are exposed to disproportional risks from superfund sites (Carter and Kalman 2020). Communities with greater proportion of minority and low-income residents experience higher health risk, because of their closer residence to superfund sites (Crawford 1994). Some reports argue that race was the most significant factor in predicting the hazardous facilities (Christ 1987). Floods exacerbate the exposure of residents to pollutants released from these industrial and toxic sites and such exposures could lead to long-term chronic health impacts. Increased flood hazards due to sea level rise can make these hazards and impacts even worse. However, limited studies exist to evaluate the extent to which current and future flood hazards disproportionately affect populations living in proximity of industrial and toxic sites.

There are also some other determinants for the threats of flood, such as water-based amenities, self-protection procedures. Usually, the regions with water-based amenities are associated with higher risk of flood, due to the abundance of water body in the amenities (Collins 2010, Collins, Grineski et al. 2018). The structures of the building, such as elevation and flood-proofing materials, will also reduce the flood impacts (Botzen, Aerts et al. 2013, de Moel, van Vliet et al. 2014). The urban built environment factors, like ground imperviousness, land use type, distance to the streamline, etc., will also affect the locations' vulnerability to flood (Mobley, Sebastian et al. 2019, Dong, Yu et al. 2020). Besides these environmental factors, another factor is the subjective human perceptions on flood, and the related information collected from social sensing (Yuan, Yang et al. 2021, Yuan, Fan et al. 2022). People's awareness of flood risk will influence selection of home locations, and flood-proof materials, which eventually affect the loss caused by the flood (Lindell and Hwang 2008, Heitz, Spaeter et al. 2009, Kellens, Zaalberg et al. 2011, Harlan, Sarango et al. 2019, Ridha, Ross et al. 2022).

# 3. Datasets and Methodology

## 3.1 Datasets

The study focuses on coastal areas in Texas as our study area. Coastal areas of Texas have a large number of industrial and toxic sites and also highly exposed to flood hazards. Four kinds of datasets are used in the study: (1) Industrial facilities, which is provided by the United States Environmental Protection Agency (USEPA) and include locations of facilities that emit air pollutants during industrial process (EPA 2022); (2) Toxic facilities, which is also provided by USEPA, and include locations of toxic facilities such as National Priority List Sites (NPL) and Toxic Release Inventory Sites (TRI) (EPA 2022); (3) flood map caused by sea level rise, which is provided by Deltares global flood map (Microsoft 2022) and provides the inundation maps of flood along coastal and the flood depth, in current and predicted future scenarios ( in the year of 2018 and 2050 separately); this open dataset flood map is one of the most reliable publicly available inundation maps available, (4) socioeconomic census data at the census tract level, provided by the United States Census Bureau (USCB). A description of the datasets is displayed in Table 1.

**Table 1**. Datasets in the study and key attributes

| Datasets | Source | Description |
| --- | --- | --- |
| Industries & Toxic facilities | USEPA | • The industrial facilities include the facilities that emit air pollutants during industrial process.<br>• The toxic facilities include the type and locations of NPL and TRI sites |
| Flood map by sea level rise in 2018 and 2050 | Deltares global flood map | The location of inundation along the coastline and the flood depth |
| Socioeconomic census data at the census tract level | USCB | Statistics data for each census tract, including total population, income, minority population, etc. |

## 3.2 Methodology

This study aims to compare the extent to which certain groups of population are threatened by flood-induced pollutant dispersion from industries and toxic facilities, under current and future flood scenarios.

Consequently, the first step is to identify the flooded facilities. We overlay the spatial distribution of industries and toxic facilities with the distribution of flood map along the coastline (as illustrated in Section 3.1). As the flood map produced by Deltares uses discrete points to represent the flood, we create a buffer of 0.1 mile based on the points in the flood map and the facilities that fall within the buffer are identified as the flooded facilities.

Then, the second step is to quantify the population that can be potentially threatened by the pollutant dispersion from the flooded facilities. To this end, we further create 1 mile, 3 miles and 5 miles buffers considering the flood impact and residents mobility patterns (Carter and Kalman 2020, Liu, Zhang et al. 2021, Liu, Wang et al. 2022),  based on the flooded facilities and the threatened population for each census tract is the tract's total population multiplied by the proportion of tract's area that falls within the flooded buffer zone. An explanation in formula is given in following equation:

$$A\_Popu_i = T\_Popu_i * \frac{Area\_in\_buffer_{i,n,m}}{T\_Area_i}$$

Where $A\_Popu_i$ is the threatened population in the $i^{th}$ census tract, $T\_Popu_i$ is the tract's total population, $Area\_in\_buffer_{i,n\_m}$ is the tract's area that falls within $n$-mile ($n$= 1, 3, and 5) buffer zone of the flooded facilities, and $T\_Area_i$ is the tract total area. We examine vulnerable populations, that are (1) below poverty, (2) of ethnical minority, (3) unemployed and (4) without high school diploma, to provide a comprehensive overview about how different communities, especially disadvantageous communities who live in proximity of industrial and toxic sites and suffer from the flooded caused by sea level rise, which put new threats on the environmental justice. The flood maps in 2018 and 2050 are used separately for comparison under current and future scenarios.

## 4. Results

### 4.1 Census tracts and population threatened by the pollutant dispersion under current scenario (2018)

Three different distances (i.e., 1 mile, 3 miles and 5 miles) are used to create buffers for the flooded facilities based on the flood map of 2018, and then identify the affected tracts (shown in Figure 1). It shows that as the distance of buffer expands, the number of increased tracts steadily increases. The north Texas along the coastline are the regions severely affected, including cities such as Houston, Beaumont, Nederland and Port Arthur.

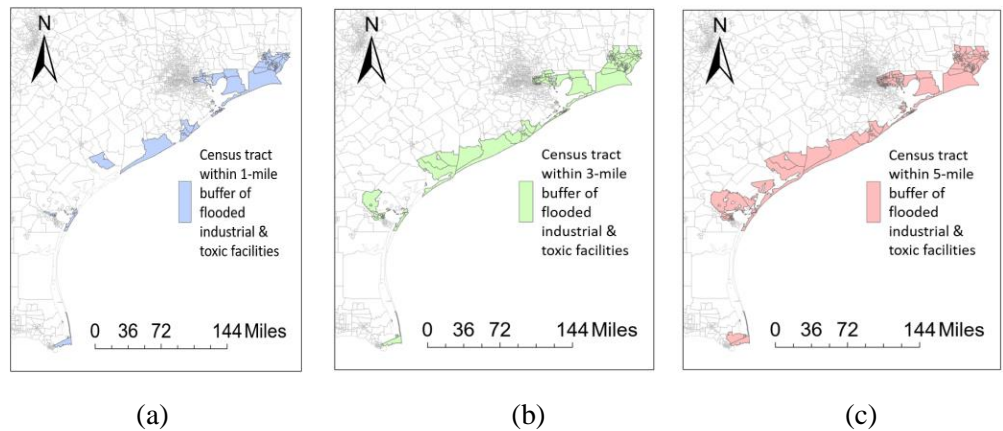

**Figure 1**. The census tracts threatened by the pollutant dispersion within (a) 1-mile buffer, (b) 3-mile buffer and (c) 5-mile buffer, in 2018.

The affected tracts along the North coastline are also the regions with higher population density and hence, smaller census tracts, compared with other regions in the central and south Texas, leading to more population being exposed to the threat of potential pollutant dispersion due to flood risk. The number of the threatened population in each census are shown in Figure 2. It can be seen that Houston and Beaumont are the regions with large amount of population threatened by the pollutant dispersion (Details in Appendix).

A similar pattern can be drawn from Figure 3, which compares the ratio of threatened population (threatened population/total population) in each census tract. The results indicate that, due to the high population density and high concentration of industrial and toxic facilities, majority of the populations along the coastline in Beaumont, Nederland and Port Arthur, are exposed to flood-induced pollutant dispersion threat. The residents in Houston and Corpus are also potentially threatened, especially along the coastal areas, where industrial and toxic facilities locate.

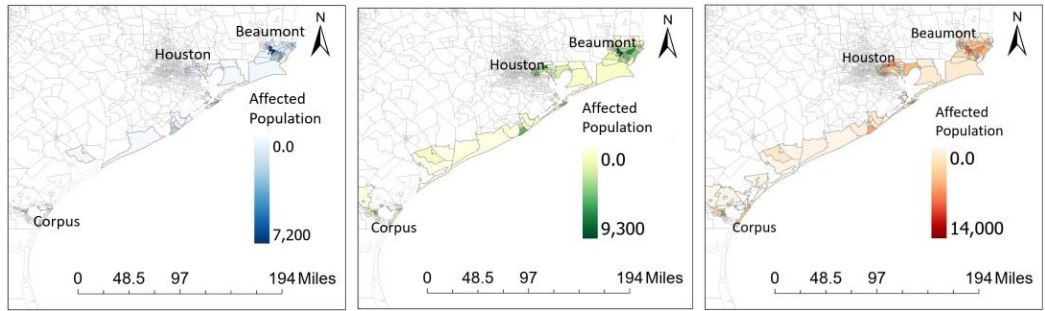

**Figure 2**. The population threatened by the pollutant dispersion in each census tract, within (a) 1-mile buffer, (b) 3-mile buffer and (c) 5-mile buffer, in 2018.

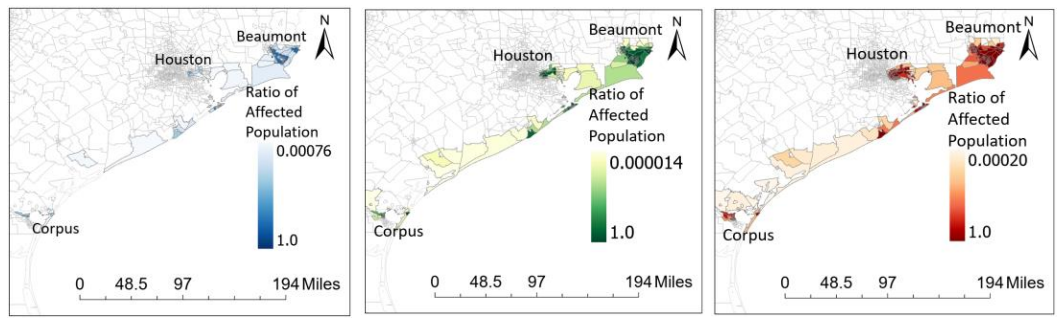

**Figure 3**. The ratio of the population threatened by the pollutant dispersion in each census tract, within (a) 1-mile buffer, (b) 3-mile buffer and (c) 5-mile buffer, in 2018.

The vulnerable communities are especially subject to the threat of potential risk. As stated in Section 3.2, this study focus on four types of socially vulnerable groups and examines the extent to which they are particularly threatened by the pollutant dispersion. Table 2 shows the mean income, proportions of minority population, below-poverty population, unemployed population and population without high-school diploma, in all the census tracts in Texas, and the tracts within certain distances of buffers of the flooded sites. A radar chart is further shown in Figure 4, with all the proportion normalized for visual interpretation. It can be clearly seen that, in the regions that are in the proximity of the flooded sites, there is a high concentration of socially vulnerable populations, whose living situation and health condition may be further exacerbated due to the potential pollutant dispersion from the flooded facilities.

Table 2. The proportions of socially vulnerable communities in different regions

|  | Mean of All Census Tracts | Mean of Tracts within 1 mile of flooded sites | Mean of Tracts within 3 mile of flooded sites | Mean of Tracts within 5 mile of flooded sites |
|---|---|---|---|---|
| Per capita income | 29,684 | 24,736 | 24,296 | 24,102 |
| Proportion of minority | 57.6% | 60.5% | 65.6% | 69.5% |
| Proportion of below-poverty population | 15.1% | 19.7% | 20.6% | 19.3% |
| Proportion of unemployed population | 2.7% | 3.1% | 3.2% | 3.3% |
| Proportion of population without high-school diploma | 10.7% | 14.0% | 16.0% | 16.0% |

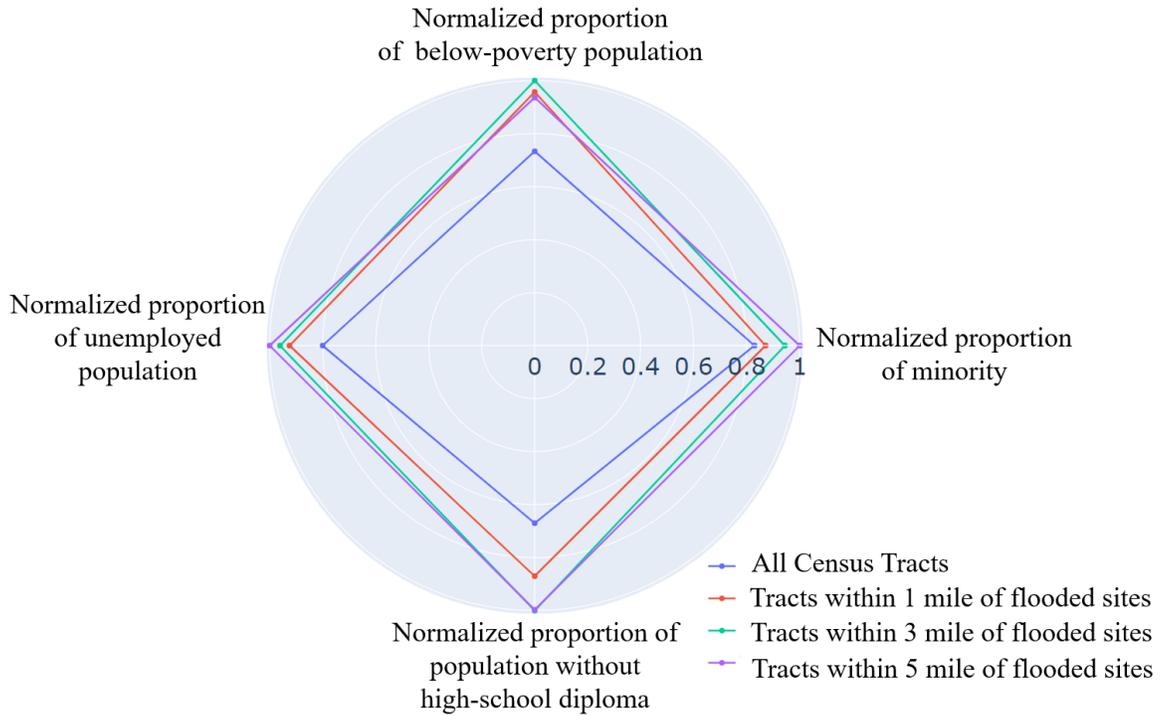

Figure 4. A radar chart of the Table 2, with the proportions normalized for visual comparison.

## 4.2 Comparison of threats by flood-induced pollutant dispersion under current (2018) vs. future (2050) scenarios

To analyze the extent to which the exposure of socially vulnerable communities may be aggravated due to the future seal level rise, the flood map in 2050 is used to identify the census tracts and population threated by the pollutant dispersion in the future, and a comparison is made between the current and future scenarios (shown in Figure 5). The results show that, compared to the current flood scenario in 2018, there will be about 10% more population threatened by flood-induced pollutant dispersion in 2050.

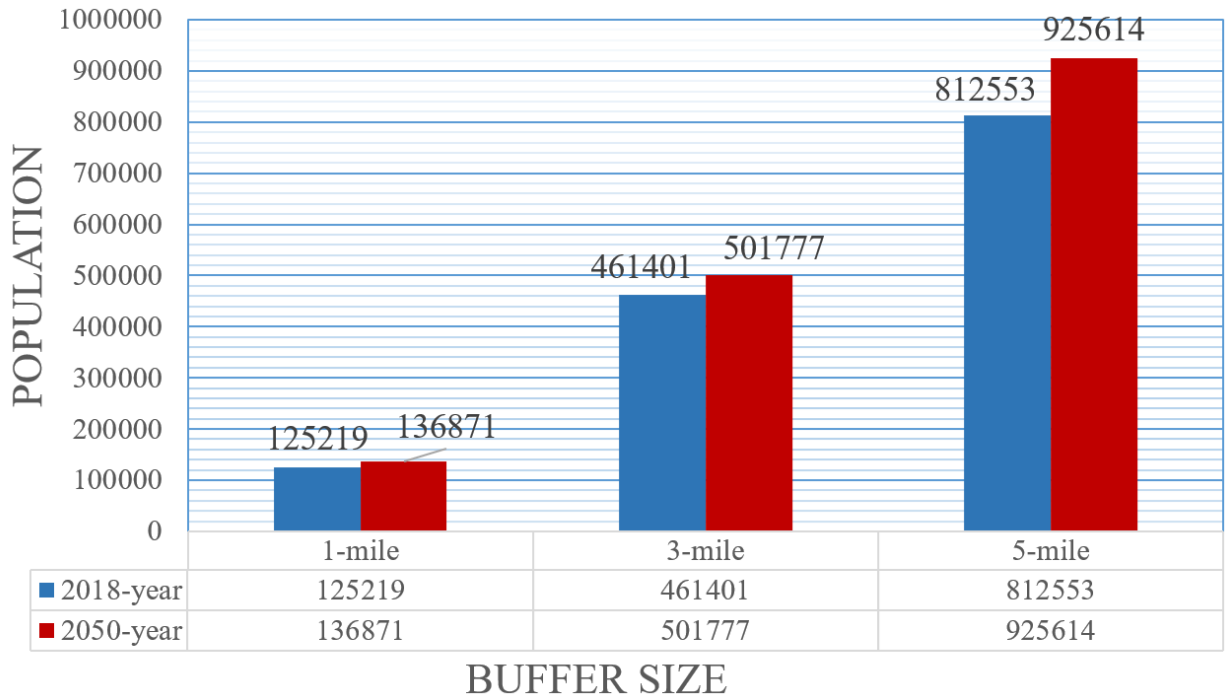

**Figure 5**. Population threatened by the flooded industrial & toxic facilities, separately in 2018 and 2050.

The increased exposure of populations that are threated in each census tract, from 2018 to 2050, is further illustrated in Figure 6. It can be seen that the sea level rise in the future will inevitably cause more population exposed to the threat of pollutant dispersion from flooded industrial and toxic facilities. The threats are especially obvious for certain regions in Houston and Beaumont. We further calculate the ratio of increased population (increased_population_under_threat/total_population) in each census tract (Figure 7). The results show that, among the cities in Texas, Houston will be the one that especially suffer from the increased flood-induced pollutant dispersion threats caused by the seal level rise. Compared with situation in 2018, certain tracts in Houston will have 50% more population affected by the pollutant dispersion in 2050 (Details in Appendix).

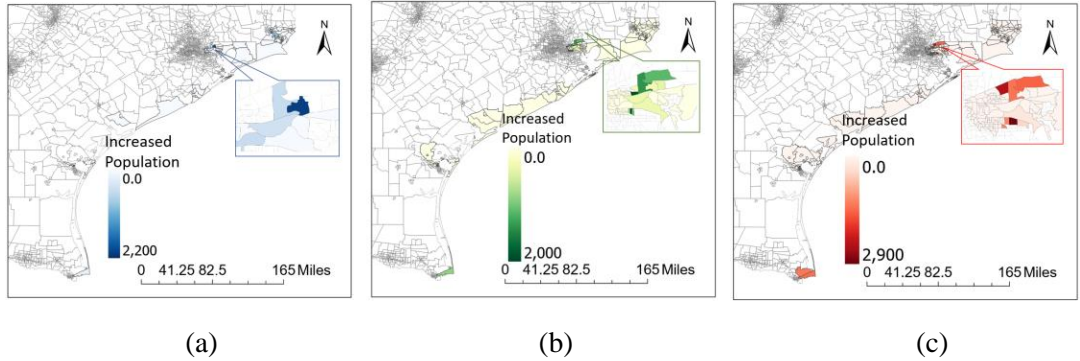

**Figure 6**. From 2018 to 2050, the increased population threatened by the pollutant dispersion in each census tract, within (a) 1-mile buffer, (b) 3-mile buffer and (c) 5-mile buffer.

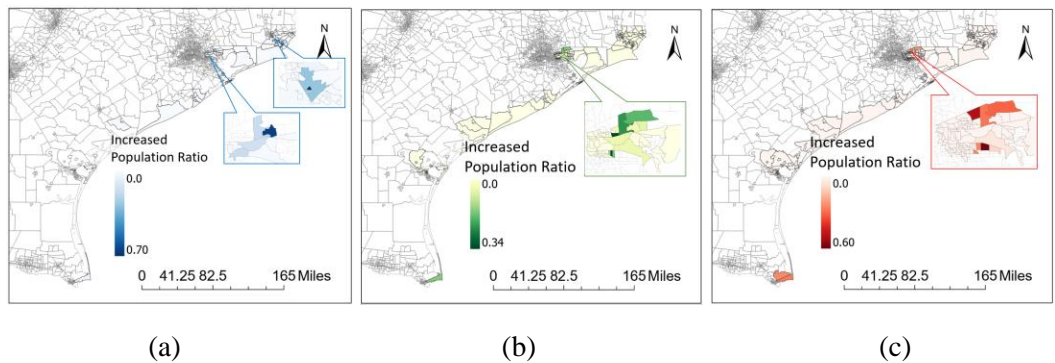

**Figure 7**. From 2018 to 2050, the ratio of increased population threatened by the pollutant dispersion to total population, in each census tract within (a) 1-mile buffer, (b) 3-mile buffer and (c) 5-mile buffer. Compared with situation in 2018, certain regions in Houston get 50% more population under flood-induced pollutant dispersion threat in 2050.

### 4.3 Who will suffer more, from the projected seal level rise?

We investigate the demographics of the additional 10% more population (Section 4.2) that will be threatened by the increased pollutant dispersion threats from the flooded facilities in 2050, and compare the results with the demographics of all the census tracts in Texas (Figure 8). The result shows that for the increased population that are under threat, the means of proportions of unemployed population (3.2% for 1-mile buffer zone, 3.3% for 3-mile buffer zone and 3.5% for 5-mile buffer zone) and population without high-school diploma (15.0% for 1-mile buffer zone, 13.2% for 3-mile buffer zone, and 13.0% for 5-mile buffer zone) are consistently higher than those of the overall population in Texas (2.7% and 10.7% respectively). On the other hand, for the census tracts that are within certain distances of the flooded facilities in 2050, the proportion of below-poverty (19.5 % for 3-mile buffer zone and 18.9 % for 5-mile zone) and minority (60.0% for 3-mile buffer zone and 63.6 % for 5-mile buffer zone) are also consistently higher than those of overall population (15.1% and 57.7%).

These results reveal that, socially vulnerable communities not only have more exposure from the potential pollutant dispersion from the flooded facilities under current flooding hazards, but also will bear a greater threat due to the increasing seal level rise in the future flood scenarios.

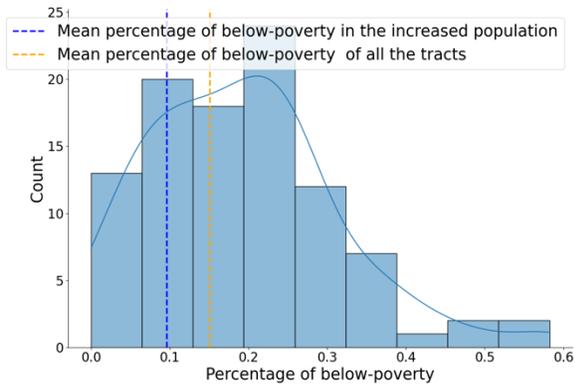
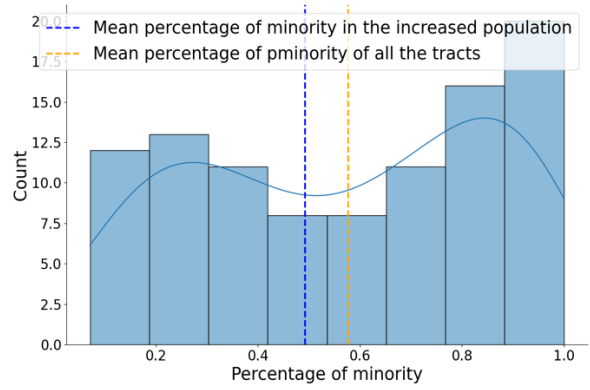
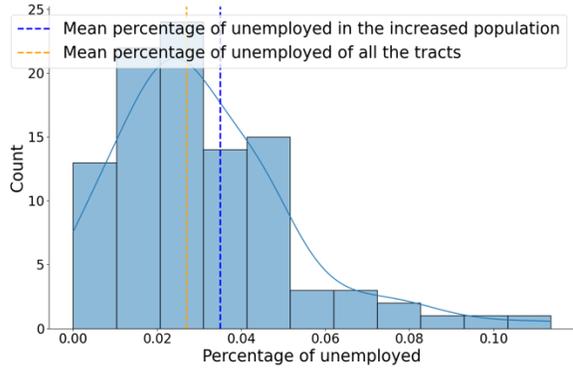
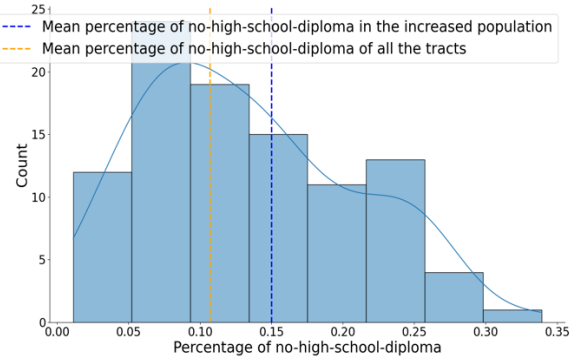

(a)

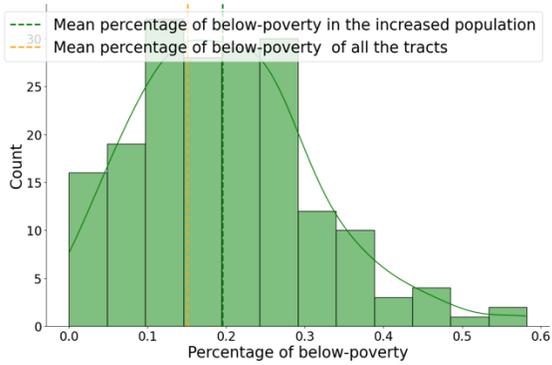
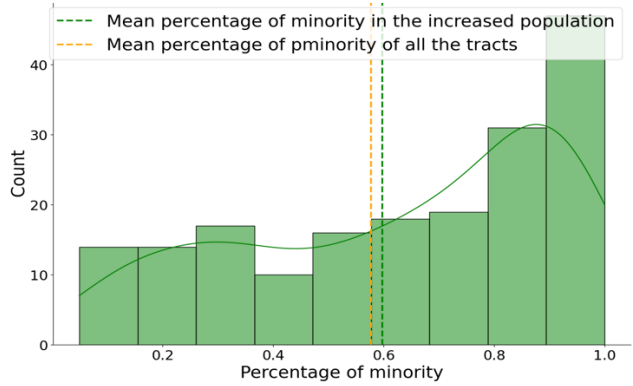
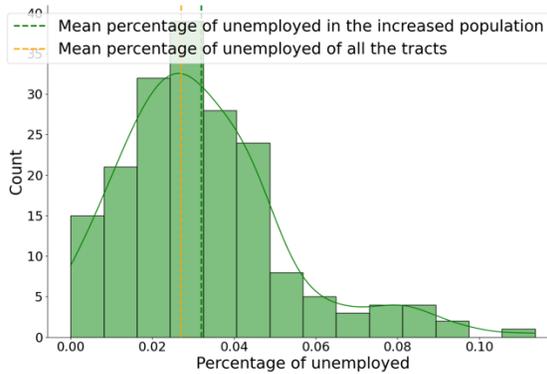
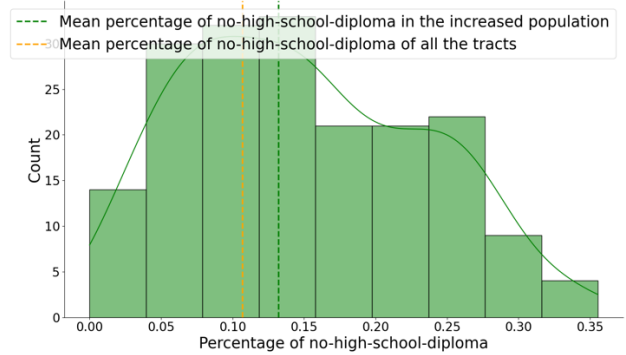

(b)

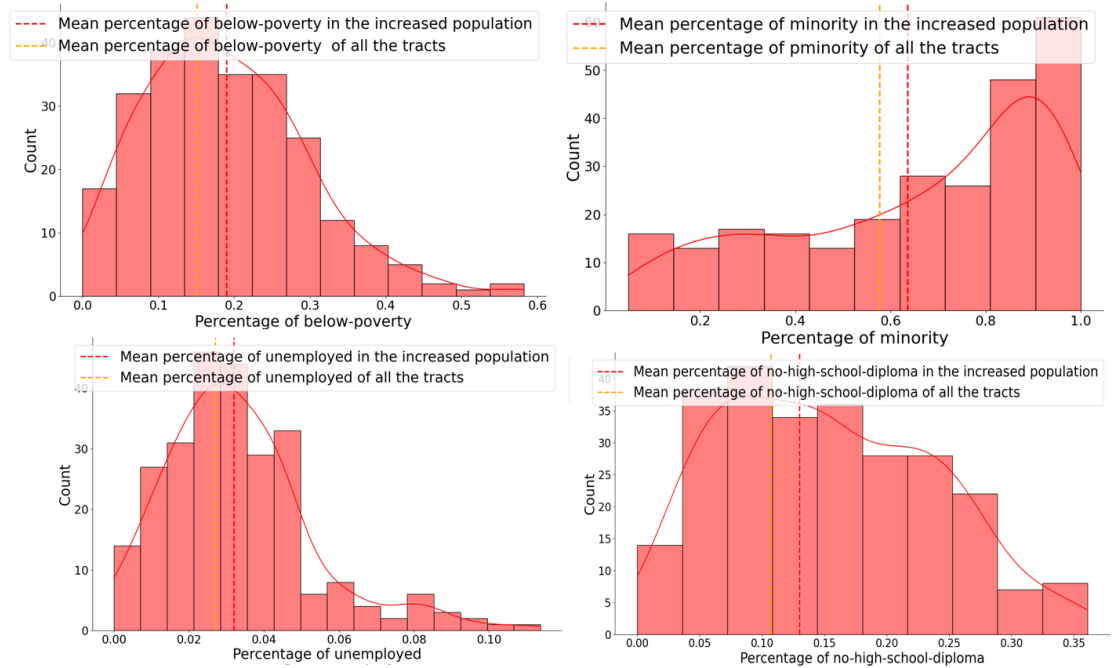

(c)

**Figure 8**. The histogram of the proportion of the (1) minority population, (2) below-poverty population, (3) unemployed population and (4) population without high school diploma among the increased population that are under threat, among the census tracts threatened by the pollutant dispersion from the flooded facilities.

## 5. Discussion
### 5.1 Regions along the north coastline in Texas are subject to higher threats
The threatened population investigated in this study are under the compound influence of the spatial distribution of population, flood hazards and industrial and toxic site facilities. For example, this study reveals that certain regions along the North coastline in Texas, such as Houston, Beaumont, Nederland and Port Arthur, are especially threatened by the pollutant dispersion from the flooded facilities. However, the underlying factors behind respective regions' threat extent are different. As shown in Figure 9, for regions in Beaumont, Nederland and Port Arthur, their threats mainly derive from the high concentration of industrial and toxic facilities along the coastline, which has put nearly all the residents in these regions under threat (Figure 3). While for Houston, the risk mainly derives from the high population density in the risky areas. Although the flooded facilities in Houston are much less compared to Beaumont, Nederland and Port Arthur, the high population density around the flooded facilities, still put the coastal areas in Houston under great threat. Such difference among different cities reveals the heterogeneous cause for the threats, underneath the seemingly similar phenomenon.

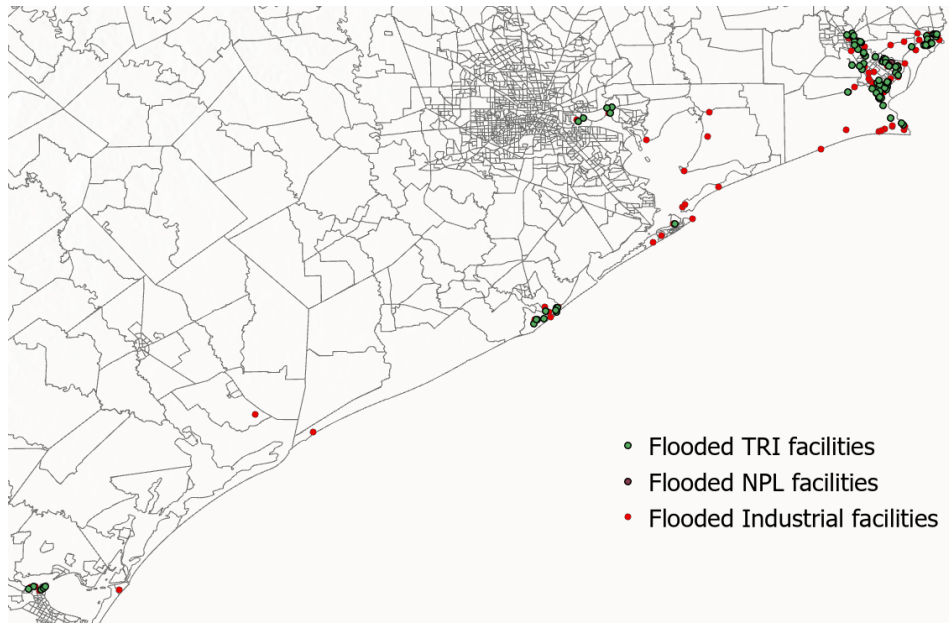

**Figure 9**. The flooded facilities in 2018. There are 230 flooded facilities in total: 113 flooded TRI facilities, 2 flooded NPL facilities, 115 flooded industrial facilities. The detailed information of the flooded sites can be found in appendix.

## 5.2 Socially vulnerable populations disproportionately suffer from the threats and will suffer more in the future

Our study draws a different conclusion from the previous studies. The previous findings conclude that non-vulnerable population are associated with higher coastal flood threats (Chakraborty, Collins et al. 2014). On the contrary, our study concludes that the socially vulnerable communities have a greater threat exposure from the coastal flood threats and the associated pollutant dispersion from flooded industrial and toxic facilities. Such difference originates from that our study defines the threat of coastal flood from a new perspective: previous works mainly identify the threat based on the 'pre-flood' risk (100-year flood zones); while our study derives the coastal flood with reference to the inundation due to the sea level rise, and identifies the threat from potential pollutant dispersion from the flooded facilities. Such perspective for viewing additional dimensions of threats of coastal flood has long been understudied by the previous works.

Contrary to the conclusions of previous works, our finding points out that, due to their social and economic status, vulnerable populations (e.g., minority, low-income/below-poverty, unemployed, without high-school diploma) are more likely to live closer to the industrial and toxic facilities, and actually suffer more from the threat of the pollutant dispersion, than other communities. Such threat is twofold, which means not only that the threatened population under current scenario have a higher proportion of socially vulnerable people, but also that, the socially vulnerable populations will still constitute a large proportion of the increased population that will be exposed to such threats in the future.

The threat of pollutant dispersion from flooded facilities may potentially put risk on the health of the disadvantageous communities, whose living and health condition will be further aggravated given their unfavorable economic status and insufficient coverage of medical insurance. These findings provide important and novel complementary insights to the existing knowledge of social inequality caused by the flooding and environmental hazards.

## 5.3 Practical implications

The threat of flood-induced pollutant dispersion, is a compound outcome affected by the spatial distribution of flood hazards, population and industrial/toxic site facilities. Hence, policies needed to alleviate the threats of flood-induced pollutant dispersion should account for the interactions among these three determinants. Based on the results from this study, we identified the following strategies that call for the actions from public officials, regulators, and decision makers to mitigate the threat from the pollutant dispersion and improve environmental justice in flood context.

The first strategy is relocating industrial and toxic facilities away the regions that are susceptible to the flood inundation along the coastline as well as populous regions. Relocation of communities is a strategy commonly discussed in view of the flood threats (Sipe and Vella 2014). The future sea level rise inevitably erodes the coastline and posits greater threat on the facilities along the coastal areas. We argue that a straightforward solution to this issue is to relocate the facilities away from the coastal areas, or compromisingly in coastal areas with less population concentrations, which is particularly important for the future practice of urban planning.

Second, it is important to emphasize the necessity for adopting measures of coastal flood protection, especially for the facilities that are infeasible to relocate within short term due to historical or economic reasons. Coastal protection measures are effective in reducing the flood threats (Hallegatte, Green et al. 2013). This may be more economically friendly than relocation of facilities, in the short term. Yet considering the long-term trend of sea level rise, the coastal flood protection may need to be strengthened or rebuilt to deal with the increasing threat, which may eventually result in larger costs. Besides, the local environmental factors such as elevation, imperviousness, distance to the streamline/coastline should be considered for customized protection that fits the local environment.

Finally, we call for greater attention to equity in policies and regulations related to environmental impacts of industrial and toxic sites, urban development plans, and flood risk reduction. Our analysis has shown that socially vulnerable populations (i.e., low income, ethnically minority, unemployed, without high-school diploma) are disproportionately exposed to a greater threat of flood-induced pollutant dispersion and the subsequent health impacts. The rising sea levels will further worsen this threat exposure, and cause more profound negative effects on the health and wellbeing of these sub-populations. Urban development plans, environmental regulations, and flood risk reduction investments should be designed and implemented in ways to equitably reduce the exposure to these threats for socially vulnerable populations (Maantay and Maroko 2009, Montgomery and Chakraborty 2015).

## 5.4 Limitation and future directions

This study and its findings contribute to a better understanding of the intersection of environmental justice issues and flood hazards under sea level rise in coastal areas. Nevertheless, the analysis in this study has some limitations and further efforts are needed for more robust results.

First, the socioeconomic census dataset used in this study can be updated in the future. Due to the data availability issue, we use the census data in current scenario to quantify the future population threatened by the pollutant dispersion from the flooded facilities. This may introduce imprecision into our results as the fluctuation of demography in the long term are not included. Future studies are needed to examine forecasted census information in examining future threats and risks to populations.

Second, the types of facilities used in this study can be expanded. Again, due to the data availability, this study includes the NPL and TRI at the toxic facilities. Some other types of facilities such as Risk Management Plan Sites (RMP) and Treatment Storage and Disposal Sites (TSD) may be further integrated into the facilities to provide a more comprehensive picture for the toxic facilities that are potentially threatened by the seal level rise.

Third, the accuracy of the flood map in this study can be refined. Deltares global flood map (Section 3.1) utilizes a series of models to produce the inundation maps of flood depth, yet lacking in consideration of certain factors such as the implementation of future coastal flood protection measures. As updated flood inundation predictions become available, similar analyses can be repeated to refresh the insights obtained in this study.

## 6. Concluding remarks

Pollutant dispersion caused by the coastal floods is an increasing risk to public health and wellbeing, especially with the growing impacts of climate change. However, most of the previous work on flood risk primarily focused on risks from floodplains, with insufficient attention to flood-induced pollutant dispersion exposure. In addition, limited number of studies have examined the effects of future sea level rise on flood-induced pollutant dispersion in coastal areas and the extent to which this risk affects the overall societal environmental justice by threating different socio-demographic subpopulations.

To address this gap, our study examined the coastal areas of Texas as a case study and utilized the flood inundation under current and future scenarios, to investigate the threats of coastal pollutant dispersion and its environmental justice implications. Our analysis shows that socially vulnerable communities are disproportionately exposed to the threats currently, and will continue to be inequitably exposed to flood-induced pollutant dispersion in the future. Specifically, socially vulnerable communities with populations of minority, low-income, unemployed, and without high-school diploma, have disparate threat exposure. To mitigate such threat exposures at the intersection of environmental injustice and sea level rise, we suggest that policies and plans should consider mitigation strategies, such as facility relocation, coastal protection, and home buyout programs. The analysis from this study provides new insights regarding ways coastal flooding endangers public health as well as our overall societal inequality. This research bridges the gap at the intersection of environmental justice and climate change impacts by investigating the extent to which pollutant dispersion due to the sea level rise threatens different communities.

**Acknowledgements**

The authors would like to acknowledge funding support from the Texas A&M X-Grant Presidential Excellence Fund.

# Appendix.

Attached as a separate file.